\documentclass[11pt,epsfig]{article}
\usepackage{graphicx}

 1
\font\elevenrm=cmr10 scaled\magstep 1

\renewenvironment{thebibliography}[1]
 { \elevenrm
   \begin{list}{\arabic{enumi}.}
    {\usecounter{enumi}     \setlength{\parsep}{0pt}
     \setlength{\itemsep}{3pt} \settowidth{\labelwidth}{#1.}
     \sloppy
    }}{\end{list}}
\begin{document}

\title{{\bf Rescaling of quantized skyrmions: \\from nucleon to baryons with heavy flavor}   \\}
\author{Vladimir B.~Kopeliovich$^{a,b}$\footnote{{\bf e-mail}: kopelio@inr.ru},
and Irina K.~Potashnikova$^d$\footnote{{\bf e-mail}: irina.potashnikova@usm.cl}
\\
\small{\em a) Institute for Nuclear Research of RAS, Moscow 117312, Russia}
\\
\small{\em b) Moscow Institute of Physics and Technology (MIPT), Dolgoprudny, Moscow district, Russia} 
\\
\small{\em d) Departamento de F\'{\i}sica, Universidad T\'ecnica Federico Santa Mar\'{\i}a;}\\
\small{\it and Centro Cient\'ifico-Tecnol\'ogico de Valpara\'iso,
Avda. Espa\~na, 1680, Valpara\'iso, Chile}
}

\date{}
\maketitle

\begin{abstract}
The role of rescaling (expansion or squeezing) of quantized skyrmions is studied for the spectrum of baryons 
beginning with nucleon and $\Delta(1232)$, and with flavors 
strangeness, charm or beauty. The expansion of skyrmions due to the centrifugal forces has influence on
the masses of baryons without flavor ($N$ and especially $\Delta$). The rescaling of skyrmions has smaller influence on the spectrum 
of strange baryons, it is more important for the case of charm, and is crucial for baryons with beauty
quantum number, where strong squeezing takes place.  Two competing tendencies are clearly observed: 
expansion of skyrmions when isospin (or spin) increases,
and squeezing with increasing mass of the flavor.
For the case of beauty baryon $\Lambda_b$ satisfactory agreement with data
can be reached for the value  $r_b= F_B/F_\pi \simeq 2.6 $, for the case of $\Sigma_b$ there
should be $r_b\sim 2$, so for the beauty flavor the method seems to be not quite satisfactory
because of certain intrinsic discrepances. Some pentaquark states with hidden strangeness, 
charm or beauty are considered as well.

\end{abstract}

\section{Introduction} 

Studies of baryons spectrum is one of important directions of the elementary particle physics, and 
the chiral soliton approach \cite{skyrme1, witten} provides one of attrctive possibilities for this, after different 
kinds of the quark models. The pioneer papers are well known \cite{anw, an}, 
where static properties of baryons (nucleon and $\Delta(1232)$) have been calculated 
 within accuracy about 30\% of experimental values. 

Two parameters of the model - pion decay constant $F_\pi$ and the Skyrme constant $e$ were defined
in \cite{anw, an} by fitting the masses of the nucleon and $\Delta(1232)$ which allowed to calculate
other static properties of these baryons. The classicaal mass of the soliton, skyrmion with baryon
(winding) number $B=1$ - was obtained by direct minimization of the static energy functional. The 
energy (mass) of the quantized states (baryons) is the sum of the classical mass and isospin 
dependent quantum correction. The chiral fields configuration for each of quantized states
does not satisfy the Euler-Lagrange equation, and the sum of the classical mass of the skyrmion
plus quantum correction could be minimized further, but this issue has not been discussed in 
\cite{anw,an} because the quantum correction turned out to be small enough for the nucleon --- less than
$\sim 10$\%, although greater and much more important for the $\Delta(1232)$, because it is 
responsible for the mass splitting between nucleon and $\Delta$.

The $SU(3)$ extention of the model has been proposed somewhat later in \cite{guad} which allowed
to calculate the mass splittings between components of the $SU(3)$ multiplets of baryons. The
$SU(3)$ violating mass term, added to the model lagrangian, was considered as a small enough 
perturbation, similar to the quantum correction in the $SU(2)$ model \cite{anw}, although in 
some cases this correction is even greater. Therefore, the problem that the energy of the
quantized state is not the minimal energy, appears here again, even in greater scale.

In present paper we investigate this problem
 using the simple variant of the quantization scheme, proposed by
Klebanov and Westerberg \cite{kwest} and slightly modified in \cite{kopzak,kopshun}.
We study the influence of the change of the scale (dimension) of the whole skyrmion on 
the energy of the quantized states with heavy flavors, and with strangeness as well. 
The squeezing of the skyrmion leads 
to considerable decrease of the energy (mass) of the quantized state,
which is especially important for the charm or beauty quantum numbers, but not so important 
for the case of strangeness. For the quantized states with $u,\,d$ flavors only (but not strangeness, 
or charm, or beauty) the expansion of the state due to the centrifugal forces takes place, instead 
of squeezing.  

Features of the chiral soliton approach are described briefly in the next section, where some static characteristics of the skyrmion
 are presented. The quantization scheme is described in section 3, where the moments of inertia of the skyrmion 
and flavor excitation energies are given as well. The spectrum of baryon states is presented in section 4. The masses of some positive parity 
pentaquarks with hidden flavor (strangeness, or charm, or beauty) are estimated in section 5 within
same approach. Final section contains discussion and some conclusions.

\section{Features of the chiral soliton approach and some static properties of the skyrmion}
The starting point of the CSA, as well as of the chiral perturbation theory, is the effective chiral lagrangian
written in terms of the chiral fields incorporated into the unitary matrix $U \in SU(2)$ in the original variant of
the model \cite{skyrme1,witten}, $U=cos\,f + isin\,f\; \vec\tau \vec n$, $n_z=cos\alpha,\,n_x=sin\alpha\,cos\beta,\,n_y=sin\alpha\;\sin\beta$,
where functions $f$ (the profile of the skyrmion), and angular functions $\alpha,\;\beta$ in general case are
the functions of 3 coordinates $x,y,z$.
To get the states with flavor $s,\; c$ or $b$ we make extension of the basic $U\in SU(2)$ to $U\in SU(3)$
with $(u,d,s)$, $(u,d,c)$ or $((u,d,b)$ degrees of freedom.

It is convenient to write the lagrangian density of the model in terms of left (or right) chiral derivative
$$ l_\mu = \partial_\mu U U^\dagger = - U\partial_\mu U^\dagger \eqno (1) $$

$$ {\cal L} = -{F_\pi^2\over 16}l_\rho l_\rho +{1\over e^2} \left[l_\rho l_\tau\right]^2+
{F_\pi^2 m_\pi^2\over 16} Tr (U + U^\dagger - 2 )  \eqno (2) $$
$F_\pi$ is the pion decay constant, its experimental value is now $F_\pi \simeq 185\,MeV$ \cite{rs};
$e$ is the constant introduced by Skyrme \cite{skyrme1}. It can be defined experimentally as well,
but the allowed interval for this parameter is wide enough presently.
Meson properties - mass, decay constants - are input of the model, and baryon properties are deduced
from meson properties, according to \cite{skyrme1, witten, anw}. This, original variant of the model \cite{skyrme1},
where soliton stabilization takes place due to the 4-th order term in the lagrangian density $(2)$, is called now the $SK4$ variant.

Mass splittings within $SU(3)$ multiplets of baryons are due to the term in the lagrangian \cite{sweig},
see also \cite{kopzak}:
$$ {\cal L}^{br}= {F_\pi^2 \widetilde m_D^2\over 24} Tr(1-\sqrt 3 \lambda_8)\left(U+U^\dagger -2\right)
+{F_D^2-F_\pi^2\over 48}Tr(1-\sqrt 3 \lambda_8)\left(Ul_\mu l^\mu + l_\mu l^\mu U^\dagger\right), \eqno (3)$$
$\lambda_8$ is the $SU(3)$ Gell-Mann matrix,  
$\widetilde m_D^2 = F_D^2m_D^2/F_\pi^2 -m_\pi^2$ includes the $SU(3)$-symmetry violation in
flavor decay constants, as well as in meson masses. $m_D$ denotes the mass of the kaon, $D$-meson or $B$-meson,
for strangeness, charm, or beauty, similar holds for $F_D$.

In this section we present some static properties of the skyrmion which are 
necessary to perform the procedure of the $SU(3)$ quantization and to obtain the spectrum of states with
definite quantum numbers. 
The quantity $\Gamma$, proportional to the sigma-term, 
$$ \Gamma (\lambda)\simeq \lambda^3{F_\pi ^2\over 2} \int (1-c_f) d^3r \eqno (4) $$
plays important role in any of known quantization modela, in rigid (or soft) rotator model, and in the 
bound state model, which simplified version we exploit here.
The scaling properties of this quantity (i.e. the behaviour under change of the dimension of
the soliton $r \to \lambda r$) are shown, which will be important in our consideration. 
Numerically, for the baryon number $B=1$ configuration, $\Gamma(\lambda =1) \sim 5\, Gev^{-1} $. 
The moments of inertia of skyrmion, the isotopical $\Theta_I \sim (5 - 6)Gev^{-1}$, the flavor $\Theta_F \sim (2 - 3)Gev^{-1}$
play an important role as well, 
see e.g. \cite{kopzak,kopshun}, and references here. All moments of inertia $\Theta \sim N_c$, where
$N_c$ is the number of colors of underlying QCD \cite{witten}. Expressions for the moments of inertia will
be given in the next section.

One of main advantages of the CSA \footnote{It is the authors opinion, probably, not accepted widely.} 
consists in the possibility to consider baryonic states with different flavors -
strange, charmed or beautiful - and with different atomic (baryon) numbers from unique point of view, using
one and the same set of the model parameters. The properties of the system are evaluated as a function
of external quantum numbers which characterize the system as a whole, whereas the hadronic content of the state 
plays a secondary role. This is in close correspondence with standard experimental situation where e.g.
in the missing mass experiments the spectrum of states is measured at fixed external quantum numbers
- strangeness or other flavor, isospin, etc. The so called deeply bound antikaon-nuclei states have been
considered from this point of view in \cite{koppot} not in condratiction with data (this is probably one of
most striking examples).

Remarkably that the moments of inertia of skyrmions carry information about their interactions. 
Probably, the first example how it works are the moments of inertia of the toroidal $B=2$ biskyrmion.
The orbital moment of inertia $\Theta_J$ is greater than the isotopic moment of inertia $\Theta_I$,
as a result, the quantized state with the isospin $I=0$ and spin $J=1$ (analogue of the deuteron) has smaller
energy than the state with $I=1,\;J=0$ (quasi-deuteron, or nucleon-nucleon scattering state), in qualitative 
agreement with experimental observation that deuteron is bound stronger.

 In the pioneer paper \cite{anw} the masses of the nucleon and $\Delta$ 
isobar have been fitted, and as a result the pion decay constant turned out to be considerably lower
than experimental value $F_\pi \simeq 185\,MeV$. Later another approach has been developed, in 
particular by Siegen University theory group (G.Holzwarth, B.Schwesinger, H.Walliser and H.Weigel).
The idea is that the value of the classical mass of the skyrmion is controlled by poorly known
loop corrections of the order of $N_c^0$, or so called Casimir energy, \cite{mouss}. Therefore, it
makes more sense to calculate the differences of masses of baryons with different quantum numbers,
like difference of masses of nucleon and $\Delta(1232)$ isobar (as it was made first in \cite{anw}),
nucleon and hyperons, i.e. mass splittings inside $SU(3)$ multiplets of baryons, calculated first 
in \cite{guad}.

The classical mass of the skyrmion is calculated usually with the pion mass term included to the 
lagrangian, and consists of 3 parts which scale differently:
$$ M_{cl} = m_1 \lambda + m_{-1}/\lambda + m_3 \lambda^3 \eqno (5) $$
with
$$ m_1= F_\pi^2{\pi\over 2}\int\left(f'^2 + 2 {s_f^2\over r^2}\right)r^2dr ,\quad 
m_{-1} = {2\pi\over e^2}\int {s_f^2 \over r^2} \left(2f'^2 + {s_f^2\over r^2}\right) r^2dr, $$
$$ m_3= \pi \,F_\pi^2 m_\pi^2 \int \left(1-c_f\right) r^2 dr ={m_\pi^2\over 2} \Gamma, \eqno(5a) $$
which satisfy the Derrick relation
$$ m_1 + 3\, m_3 = m_{-1} $$
see table 1.
\section{Rigid oscillator quantization model, moments of inertia of the skyrmion}
We shall use the following mass formula for the quantized state derived in \cite{kwest} for
the quantization scheme used here
$$M(B=1,F,I,J) = M_{cl}+ |F|\omega_{F} + \Delta E_{1/N_c}. \eqno (6) $$

The flavor (antiflavor) excitation energies are
$$\omega_{F} = {3\over 8\Theta_{F}} (\mu_{F} -1 ); \quad \bar\omega_{F} = {3\over 8\Theta_{F}} (\mu_{F} +1 ) \eqno(7) $$
with
$$\mu_{F}=\left[1+{16 [\bar m_D^2\Gamma +(F_D^2-F_\pi^2)\widetilde \Gamma]\Theta_{F}\over 9}\right]^{1/2} \eqno(8)$$

$$ \widetilde \Gamma = {1\over 4}\int c_f\left[c_f (\vec\partial f)^2 + s_f^2(\vec \partial n_i)^2 \right] d^3r, \eqno(9)$$
see \cite{kopzak,kopshun}. Evidently, $ \widetilde \Gamma \sim \lambda$ under scaling procedure.

Different terms in $(6)$ scale differently as the number of colors in this expression:
$$M_{cl} \sim N_c, \quad \omega_F \sim N_c^0, $$ all moments of inertia $\Theta \sim N_c$. 

Previously estimates of the flavor excitation energies were made mostly in perturbation theory,
i.e. the flavor excitation energy has been simply added to the skyrmion energy. This is not justified,
however, when the flavor excitation energy is large. Here we include this energy into simplified minimization
procedure which is made by means of the change of the soliton dimension (rescaling of the soliton).
This procedure takes into account the main degree of freedom of the $B=1$ skyrmion (hedgehog), and 
skyrmions given by the rational map anzatz \cite{hms} which has been applied successfully to describe some properties
of nuclei. 
As we show here, the rescaling leads to considerable decrease of the energy of states, beginning with the $\Delta(1232)$.
Similar (although not the same) modification of the quantized skyrmion was made, in particular, by B.Schwesinger et al \cite{kss} to improve
the description of strange dibaryon configurations.

The hyperfine splitting correction to the energy of states which is formally of the $1/N_c$ order
in the number of colors, has been obtained previously in \cite{kwest} and reproduced in \cite{kopzak,kopshun}:
$$\Delta E_{1/N_c} ={1\over 2\Theta_I}\left[c_FI_r(I_r+1) +(1-c_F) I(I+1) +
(\bar c_F-c_F) I_F(I_F+1)\right] \eqno(10) $$

$I$ is the isospin of the state,
$I_F$ is the isospin carried by flavored meson $(K,\,D,$ or $B-$meson, for unit 
flavor $I_F=1/2$),  $I_r$ can be 
interpreted as "right" isospin, or isospin of basic non-flavored configuration.
The hyperfine splitting constants
$$c_F= 1 -  {\Theta_I(\mu_F - 1) \over 2\Theta_F\mu_F}, \quad  
\bar c_F= 1 -  {\Theta_I(\mu_F - 1) \over \Theta_F\mu_F^2}, \eqno(11)$$

This correction is considered usually as small one, but it should be
included into the minimization procedure, when isospin $I$ is large.  Here we include this correction
to the masses for all baryons.

 At large enough $m_D$ the expansion can be made
$$\mu_{F} \simeq {4\bar m_D(\Gamma \Theta_{F})^{1/2}\over 3}+
{3\over 8\bar m_D\Gamma \Theta_{F}}, $$
therefore 
$$\omega_{F} \simeq {1\over 2}\bar m_D
\left({\Gamma\over \Theta_{F}}\right)^{1/2}
 -{3\over 8\Theta_{F}}. \eqno(12) $$

Here we take the ratio of decay constants $F_K/F_\pi \simeq 1.197 $,  $F_D/F_\pi \simeq 1.58$
according to the analysis performed by Rosner and Stone in \cite{rs}.
Our results presented in this paper suggest that the ratio $F_B/F_\pi$ should be greater, between
$2.$ and $2.6$.

The flavored moment of inertia equals  
(we added the rescaling factor - some power of the parameter $\lambda $ to make evident the behaviour
under the rescaling procedure $r\to r\lambda$)

$$ \Theta_F = \lambda f_1 + \lambda^3 f_3^{(0)} {F_D^2 \over F_\pi^2} = \Theta_F^{(0)} + 
\lambda^3 f_3^{(0)} \left({F_D^2 \over F_\pi^2} -1\right) \eqno (13) $$
with

$$ f_1 ={\pi\over 2 e^2} \int (1-c_F)\biggl(f'^2 +2 {s_f^2\over r^2}\biggr) r^2 dr; \qquad
 f_3^{(0)} ={\pi \over 2} F_\pi^2 \int (1- c_F) r^2 dr \eqno (14)$$
In the integrands $f$ denotes the profile function of the soliton (skyrmion).
Here we show explicitly the dependence of different parts of the inertia on the rescaling parameter $\lambda$.
In table 1 we present numerical values for  $f_1,\;f_3,\;t_1, t_3$ and other quantities used to
perform calculations of the masses of quantized states.

There is simple connection between total moment of inertia in the $SK4$ variant of the model, the
$\Theta_F $ and the sigma-term:
$$ \Theta_F^{tot} = {F_D^2\over 4F_\pi^2}\Gamma + \Theta_F = {F_D^2\over F_\pi^2}f_3^{(0)} + f_1. \eqno(15) $$

Similarly, the isotopic moment of inertia $\Theta_I$ within the rational map aproximation can be written as

$$ \Theta_I = \lambda \,t_1 + \lambda^3 \,t_3  \eqno (16) $$
with
$$ t_1 ={4\pi\over 3} \int  {2\,s_f^2 \over e^2}\left(f'^2 +B{s_f^2\over r^2}\right) r^2 dr, \qquad t_3 ={2\pi\over 3} F_\pi^2 \int s_f^2  r^2 dr. \eqno (17) $$
\newpage
\begin{center}
\begin{tabular}{|l|l|l|l|l|l|l|l|l|l|l|}
\hline
$$& $t_1$&$f_1$&$t_3$&$f_3$&$\Gamma$&$\widetilde \Gamma$&$m_1$&$m_{-1} $&$m_3$\\
\hline
$ANW$ & $3.21$  & $-$    &$1.90$      &$-$ & $-$&$-$&$432$ & $432$ & $ 0$\\ \hline
$AN $ & $3.21$  & $-$    &$1.90$      &$-$ & $-$&$-$&$357$ & $471$ & $38$\\ \hline
$Siegen$ & $3.49$  & $0.83$    &$2.07$      &$1.20$ & $4.80$&$15.6$&$759$ & $897$ & $46$\\ \hline
\end{tabular}
\end{center}
{\bf Table 1.} 
Numerical values of the quantities used in present paper, which have definite scaling
behaviour. $t_1,\,t_3,\,f_1,\,f_3,\,\Gamma,\, \widetilde \Gamma$ are in $GeV^{-1}$, $m_1,\,m_2$ and
$m_3$ are in $MeV$. First line corresponds to original parametrization of \cite{anw}, $F_\pi=129 \,MeV,\; e=5.45$
\footnote{We used the values of
the soliton mass and the moment of inertia given in \cite{anw} by formulas after Eq. (9).}, 2-d line corresponds 
to the parametrization of \cite{an}, $F_\pi=108\,MeV,\; e=4.84$, 3-d line ---
to parametrization of Siegen University group with $F_\pi = 186\,MeV$,  $e=4.12$.  \\

In the $SK6$ variant of the model the skyrmion stabilization takes place due to the 6-th order term (in chiral derivatives) in the lagrangian density,
which is proportional to the baryon number density squared, see e.g. \cite{kjp}. We shall consider this variant of the model elsewhere.


\section{Rescaling of the lowest baryons masses} 

 We are using the following expressions for the masses of baryons:

$$M_N = M_{cl} + {3\over 8\Theta_I};\quad M_\Delta = M_{cl} + {15\over 8\Theta_I}; $$

$$M_\Lambda = M_{cl} + {3\over 8\Theta_I} + \omega_F -{3(\mu_F - 1)\over 8\mu_F^2 \Theta_F};\quad
M_\Sigma = M_{cl} + {3\over 8\Theta_I} + \omega_F -{3(\mu_F - 1)\over 8\mu_F^2 \Theta_F} +
{\mu_F-1 \over 2\mu_F\Theta_F}. \eqno (19) $$

The flavor inertia $\Theta_F^{(0)}$ is the same for all three flavors, see Eq. (13), but the quantity $\mu_F$ 
and the flavor excitation energy $\omega_F$ are different for different flavors. 

Description of the masses of strange hyperons $\Lambda_s$ and $\Sigma_s$ is not perfect in table 2, because the 
configuration mixing, i.e. the mixing between the states with same isospin and strangeness, but which 
belong to different $SU(3)$ multiplets, is not taken into account in our approach. 
Moreover, after rescaling these states cannot mix, because they have different properties 
(dimensions, in particular). Satisfactory agreement with data has been obtained in \cite{sweig} just due to 
including such mixing into consideration. Improvement of the fit is certainly possible in our case
as well, by changing of the model parameters, first of all.

For the case of the non-flavored baryons the expansion of the configuration takes place due to
centrifugal forces. Technically it appears from the contribution of the spin-isospin dependent
term in the energy of quantized state, which contains the isotopic (or orbital) inertia in the 
denominator. Both isotopic and orbital inertia are proportional to the scale factor $\lambda$,
or $\lambda^3$, and increase of $\lambda$ (expansion of the skyrmion) leads to the decrease of
energy.

\begin{center}
\begin{tabular}{|l|l|l|l|l|l|l|}
\hline
$$& $\lambda_{min} $&$\delta M$&$M_B\,-\,M_N$&$(M_B\,-\,M_N)_{exp}$               \\  \hline
$\;N_{ANW}$         &     $1.1343$ & $8.1$ & $\quad - $  & $\qquad - $                 \\ \hline
$\Delta(1232)_{ANW}$  &  $1.4982$ & $128$ &$\quad 174$  & $\qquad 293$                \\ \hline
$\;\;N_{AN}$         &     $1.0978$ & $5.2$ & $\quad - $  & $\qquad - $                 \\ \hline
$\Delta(1232)_{AN}$  &  $1.3559$ & $199$ &$\quad 182$  & $\qquad 293$                \\ \hline
\hline
$N $         &     $1.0539$ & $3.1$ & $\quad - $  & $\qquad - $                 \\ \hline
$\Delta(1232)$  &  $1.2247$ & $61$ &$\quad 212$  & $\qquad 293$                \\ \hline
$\Lambda_s $ &$0.8499$ & $26 $  & $\quad 247$   & $\qquad 177 $     \\ \hline
                                            
$\Lambda_c $        & $0.5478$ & $136$ & $\quad 1432$      & $\qquad 1347$      \\ \hline
$\Lambda_b(r=2.60)$ & $0.2212$ & $2827$ & $\quad 4663 $          & $\qquad 4680$  \\ \hline
$\Lambda_b(r=2.65)$ & $0.2192 $ & $2855$ & $\quad 4724$        & $\qquad 4680$   \\ \hline
$\Sigma_s $ & $1.0221$ & $0.4$     & $\quad 370$            & $\qquad 251 $     \\ \hline
$\Sigma_c $       & $0.8920$ & $10.$       & $\quad 1682$            & $\qquad 1515$  \\ \hline

$\Sigma_b(r=2.00) $ & $0.2551 $ & $\;995 $ & $\quad 4962$          & $\qquad 4874$     \\ \hline
$\Sigma_b(r_b=2.05) $ & $0.2512 $ & $1057 $ & $\quad 5049$        & $\qquad 4874$     \\ \hline
\end{tabular}
\end{center}

{\bf Table 2.}  The values of $\lambda_{min}$ at the minimal total energy (mass) of 
quantized baryon states. The decrease of masses $\delta M$ due to the change of $\lambda$
from $1$ to $\lambda_{min}$ is given in $MeV$. The difference of masses
$M_B - M_N$, theoretrical and experimental values, are presented as well (in $MeV)$.
The first 2 lines correspond to the parametrization in the chiral symmetry limit, proposed in \cite{anw}.
Lines 3-4 correspond to the parametrization of \cite{an}, the pion mass included into lagrangian.
Other calculations were performed taking Siegen parametrization, $F_\pi = 186\,MeV,\; e=4.12$. \\

The flavor excitation energies are proportional to the mass $\bar m_D$, which is large for charm or beauty,
and to $\sqrt\Gamma \sim \lambda^{3/2}$, and this explains, why $\lambda_{min}$ is so small for beauty.

\section{Estimates of the masses of pentaquarks with hidden flavor}
For the case of pentaquarks with hidden flavor, i.e. containing  the pair of quark and antiquark, or
the pair of $D$ and $\bar D$ (or $K$ and $\bar K$, or $B$ and $\bar B$) mesons, we take in Eq. $(10)$
$I_r = I$ and $I_F=0$, and come to the energy (mass) of the state 
$$ M_{P_F} = M_{cl} + {3\mu_F\over 4\Theta_F} +  {I(I+1)\over 2\Theta_I}. \eqno (20) $$
which we minimize numerically.
Some results are shown in table 3.

Tables 2 and 3 illustrate well two competing tendencies for quantized skyrmion states:
the squeezing with increazing flavor excitation energy, and expansion due to
centrifugal forces which becomes stronger with increasing spin (isospin). For beauty
squeezing dominates in all cases considered here.

\begin{center}
\begin{tabular}{|l|l|l|l|l|l|l|}
\hline
$B(I=J)$& $\lambda_{min} $&$\delta M$&$M_B\,-\,M_N$  \\  \hline
$P_s(1/2) $         & $1.05771$ & $4.8$ & $\quad 956 $      \\ \hline
$P_s(3/2) $         & $1.2006$ & $58 $ & $\quad 1173 $      \\ \hline
$P_s(5/2) $         & $1.3884$ & $231 $ & $\quad 1449 $      \\ \hline
\hline
$P_c(1/2) $         & $0.7235$ & $86 $ & $\quad 3257 $     \\ \hline  
$P_c(3/2) $         & $0.9808$ & $0.4 $ & $\quad 3612 $      \\ \hline
$P_c(5/2) $         & $1.2700$ & $90$ & $\quad 3972 $      \\ \hline
\hline
$P_b(1/2,r_b=2.0) $         & $0.2798$ & $3828$ & $\quad 9073 $      \\ \hline
$P_b(1/2,r_b=2.1) $         & $0.2735$ & $3960$ & $\quad 9341 $      \\ \hline

$P_b(3/2,r_b=2.0) $         & $0.3223$ & $2139$ & $\quad 10432 $      \\ \hline
$P_b(3/2,r_b=2.1) $         & $0.3150$ & $2837$ & $\quad 10734 $      \\ \hline

$P_b(5/2,r_b=2.0) $ & $0.4099$ & $1359$ &$\quad 12267 $       \\ \hline
$P_b(5/2,r_b=2.1) $ & $0.4008$ & $1404$ &$\quad 13556 $       \\ \hline

\end{tabular}
\end{center}

{\bf Table 3.}  The values of $\lambda_{min}$ at the minimal total energy (mass) of 
some pentaquark states. The decrease of masses $\delta M$ due to the change of $\lambda$
from $1$ to $\lambda_{min}$ is given in $MeV$, similar to table 2. The difference of masses
$M_B - M_N$, theoretrical values only, are presented as well (in $MeV$).
Calculations are performed taking parametrization $F_\pi = 186\,MeV,\; e=4.12$.\\

The states considered in table 3 have positive parity, as a consequence of the quantization scheme used, and
isospin which coincides with the right isospin and equals to spin of the state - because the quantized
configuration of fields is of hedgehog type.
Pentaquarks with $I=J=1/2$ could belong to the antidecuplet of corresponding $SU(3)$ group, $(p,q)=(0,3)$,
those with $I=j=3/2$ could belong to the $\{27\}$-plet, $(p,q)=(2,2)$ and 
pentaquarks with $I=J=5/2$ could belong to the $\{35\}$-plet with $(p,q)=(4,1)$.

To obtain the masses of pentaquarks predicted by this simplified model, one should add the nucleon mass,
$939\,MeV$, to the numbers of the last column of table 3.
The hidden strangeness pentaquark states presnted in table 3 have masses by few hundreds of $MeV$ greater
than such states discussed previously in connection with the low-lying positive strangeness pentaquark
$\Theta^+ (1540)$, see e.g. discussion in \cite{vkip}. For example, $M[P_s(J=1/2)] = 1895\,MeV$.
The hidden charm pentaquark state has the mass by $\sim 100 MeV$ greater than the mass of the state observed
by the LHCb collaboration, $M(P_c)\simeq 4450\,MeV$:  $M[P_c(J=3/2)] \simeq 4550\,MeV$.
\section{Conclusions and prospects}
We have demonstrated that considerable decrease of the energy of quantized skyrmion states (baryons) takes place
due to change of the skyrmton dimension (rescaling). Even for baryons with $(u,d)$ flavors, nucleon and 
$\Delta(1232)$ isobar, the expansion of the skyrmion due to centrifugal force decreases the mass splitting
between $N$ and $\Delta$ considerably, and destroys the fit of masses made in \cite{anw} and \cite{an} \footnote{It is not difficult
algebraic work to define the parameters of the model $F_\pi$ and $e$ in the chiral symmetry limit of \cite{anw}.
It has been obtained in \cite{anw} for the soliton mass $M_{cl} = a F_\pi/e = (5M_N-M_\Delta)/4$, and for the mass splitting between
$\Delta (1232)$ and nucleon $\Delta_M=M_\Delta - M_N = b F_\pi e^3$, $a=36.5$,  the constant $b$ can be extracted from relation for the moment of 
inertia presented in \cite{anw} after Eq. $(9)$. 
It follows then immediately, $F_\pi =[M_{cl}^3\Delta_M/(a^3b)]^{1/4}$
and  $e=[a\Delta_M/(b M_{cl})]^{1/4}$. After rescaling there are no such simple relations, but $F_\pi$ and $e$ should be 
somewhat greater, $F_\pi \sim 150 \,MeV$.}.
This fit could be recovered by some increase of parameters of the model - towards better agreement with data, 
which demands certain technical work.

The change of the skyrmion dimensions leads to considerable lowering of the
energy (mass) of the quantized states with quantum numbers charm or beauty. For strangeness the 
effect takes place as well, but not so important - small enough in most of cases. In our estimates 
we used simple and transparent variant of the quantization procedure, originally proposed 
in \cite{kwest} and modified later in \cite{kopshun}. It is quite obvious that the
effect is important in any variant of the quantization scheme. There are two competing tendencies,
to decrease the dimension of the skyrmion when the flavor excitation energy becomes large, and to 
expand the skyrmion due to centrifugal forces, when the spin (or isospin) becomes greater.
For large enough spin (isospin) of the state the expansion takes place due to centrifugal force - 
instead of squeezing. 
For strangeness this effect dominates already for $\Sigma_s$ hyperon, see table 2, and takes place 
for all hidden strangeness pentaquarks, and for hidden charm pentaquark with $J=5/2$ (table 3).

There is some discrepance in description of masses of beauty baryons $\Lambda_b$ and $\Sigma_b$, which
 indicates that method itself is limited in its applicability. To describe the mass of
the $\Lambda_b$ baryon the ratio $r_b = F_B/F_\pi$ should be about $2.6$ ,
but to obtain satisfactory description of the mass of the $\Sigma_b$ baryons there should be $r_b \sim 2.0$. 

Modifications and improvemants of the approach, including its fine tuning, seem to be of interest. \\

{\bf Acknowledgements.}
We thank Yura Ivanov and Elena Tourdakina for help in numerical computations.
This work was supported in part by Fondecyt (Chile) grant 1130549,
by Proyecto Basal FB 0821 (Chile), and by CONICYT grant  PIA ACT1406 (Chile).
\\


{\bf References}

\baselineskip=12pt
\begin{thebibliography}{99}


\bibitem{skyrme1} T.H.R.~Skyrme, Proc.R.Soc. London, A260 (1961) 127; Nucl.Phys. 31 (1962) 556
\bibitem{witten} E.~Witten, Nucl.Phys. B223 (1983) 422; ibid B223 (1983) 433  
\bibitem{anw} G.~Adkins, C.~Nappi, and E.~Witten, Nucl.Phys. B228 (1983) 552

\bibitem{an} G.~Adkins and C.~Nappi, Nucl.Phys. B233 (1984) 109

\bibitem{guad} E.~Guadagnini, Nucl. Phys. B236 (1984) 35

\bibitem{kwest} K.M.~Westerberg and I.R.~Klebanov, Phys. Rev. D50 (1994) 5834; Phys. Rev. D53
(1996) 2804

\bibitem{kopzak} V.B.~Kopeliovich and W.J.~Zakrzewski, JETP Lett. 69 (1999) 721 [Pisma Zh.Eksp.Teor.Fiz. 69 (1999) 675];
 Eur.Phys.J. C18 (2000) 369 

\bibitem{kopshun} V.B.~Kopeliovich and A.M.~Shunderuk, J.Exp.Theor.Phys. 100 (2005) 929 
[Zh.Eksp.Teor.Fiz. 127 (2005) 1055]

\bibitem{rs} J.L.~Rosner and S.~Stone. Leptonic Decays of Charged Pseudoscalar Mesons.
arXiv:1002.1655 [hep-ex] 

\bibitem{sweig} B.~Schwesinger and H.~Weigel, Phys. Lett.B267 (1991) 438;  Nucl.Phys. A540 (1992) 461

\bibitem{koppot} V.B.~Kopeliovich and I.K.~Potashnikova,   Phys.Rev. C83 (2011) 064302 

\bibitem{mouss} B.~Moussallam,  Annals Phys. 225 (1993) 264; B.~Moussallam and D.~Kalafatis, Phys. Lett. 
272 B (1991) 196

\bibitem{hms} C.~Houghton, N.~Manton, and P.~Sutcliffe, Nucl. Phys. B 510 (1998) 507

\bibitem{kss} V.B.~Kopeliovich, B.~Schwesinger, and B.E.~Stern, Phys. Lett. B242 (1990) 145	
 
\bibitem{kjp} V.B.~Kopeliovich, J. Phys. G: Nucl. Part. Phys. 28 (2002) 103

\bibitem{vkip} V.B.~Kopeliovich and I.K.~Potashnikova, Phys.Rev. D93 (2016) no.7, 074012 





\end{thebibliography}
\end{document}